# Performance Analysis and Design of an Optomechanical Microphone Using an Integrated Photonic Waveguide Interferometer

**Xiaoyu Niu, Yuqi Meng, Zihuan Liu, Ehsan Vatankhah, Neal Hall**

Department of Electrical and Computer Engineering, The University of Texas at Austin, Austin, Texas, 78712, USA; xyniu@utexas.edu; yuqimeng@utexas.edu; zihuanliu@utexas.edu; e.vatankhah@utexas.edu; nahall@utexas.edu

## I.ABSTRACT

We present an optomechanical microphone based on a diaphragm-integrated photonic waveguide Mach–Zehnder interferometer. Acoustic pressure deforms the MEMS diaphragm, inducing strain in the sensing waveguide and changing its optical path length. We analytically evaluate the optical and mechanical transduction mechanisms and key figures of merit, including signal-to-noise ratio, dynamic range, acoustic overload pressure, and minimum detectable pressure. Two design cases are considered: a MEMS microphone and a measurement microphone. The results indicate competitive performance but no substantial overall advantage over state-of-the-art microphones in conventional applications. The architecture may nevertheless offer advantages for high-temperature and other harsh-environment sensing applications.

## II.INTRODUCTION

An optical interferometer has been known as an impactful scientific instrument for more than two centuries.[1] The most historically significant applications are the Michelson-Morley experiment[2] and the Laser Interferometer Gravitational-Wave Observatory (LIGO)[3], shown in Figure 1(a) and (b), respectively. The former demonstrated the nonexistence of “luminiferous aether.” The latter detected cosmic gravitational waves from the powerful collision of two black holes. With the recent development of photonic integrated circuits, optical interferometers enable various sensing applications, such as biomedical sensors[4], angular velocity sensors[5,6], chemical sensors[7], and accelerometers[8]. Related integrated-photonic sensors include our cantilever-based Mach–Zehnder accelerometer and our silicon-nitride accelerometer with heterodyne detection, in which deformation of a waveguide-bearing mechanical structure modulates optical phase.[20,21]

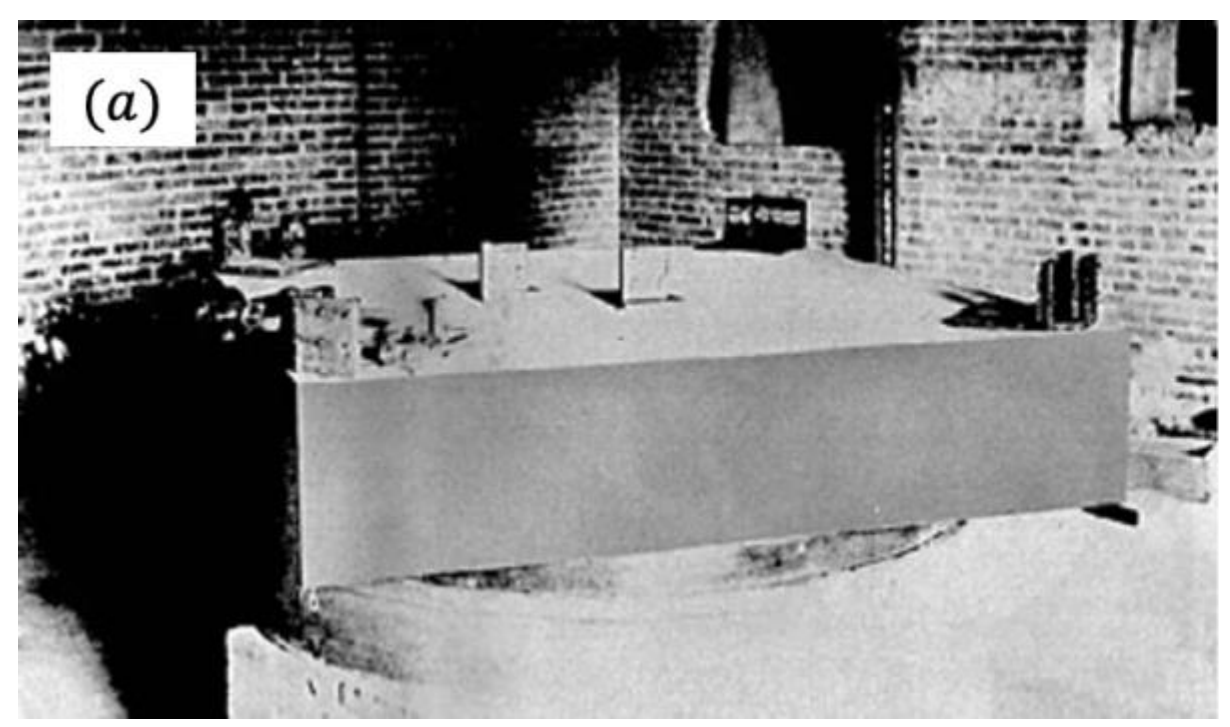

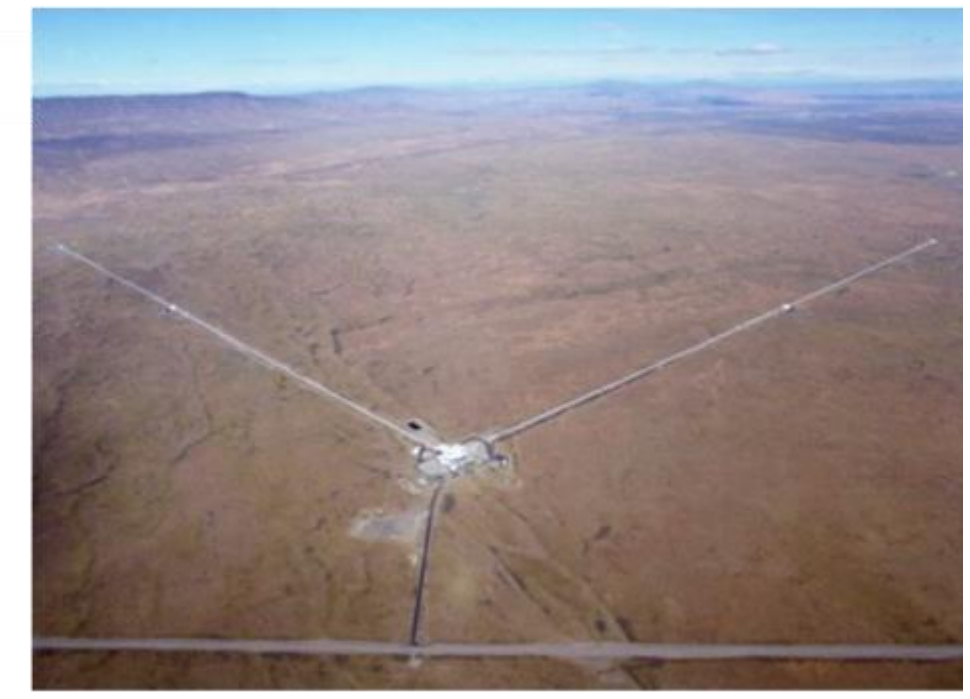

***Figure 1(a) Michelson and Morley’s interferometric setup in 1887, mounted on a stone slab that floats in an annular trough of mercury. (b) The Laser Interferometer Gravitational Wave Observatory (LIGO) near Hanford, Washington, U.S.***

Using a photonic waveguide interferometer, we designed an optomechanical microphone. Most commercial microphones rely on capacitive[9] or piezoelectric[10] readout of a vibrating diaphragm. Optical methods have also been used to read the motion of a vibrating microphone diaphragm. Most optical microphone demonstrations to date use free-space optics. We designed an optical interferometric microphone that uses a photonic waveguide embedded within the diaphragm. The deflection of the moving diaphragm generates a strain field within the diaphragm, and this strain in turn changes the optical path length of the waveguide. Light traversing the sensing waveguides of the interferometer combines with light traversing an on-chip reference waveguide to yield an optical output power that is proportional to the instantaneous displacement of the diaphragm. Researchers have demonstrated optical waveguide microphones[11–13], but their figures of merit have not been carefully discussed. We analyzed the figures of merit of an optical waveguide microphone from the perspective of MEMS microphones. We presented a detailed analysis of the signal-to-noise ratio (SNR), acoustic overload pressure (AOP), and dynamic range (DR). We explored the design for two application cases using an optical waveguide interferometer: one is a MEMS microphone; the other is a measurement microphone. The diaphragm-integrated waveguide concept considered here was first described in our 2023 conference abstract.[22]

The present waveguide-based approach differs from both electrical diaphragm readout and free-space optical interrogation. For comparison, our group has demonstrated piezoelectric sensing in a bimorph lithium niobate microphone[23] and used a MEMS microphone architecture in reverse as an air-coupled electrostatic ultrasound transmitter.[24] We have also measured airborne ultrasound with laser Doppler vibrometry by sensing sound-induced refractive-index variations along a free-space optical path.[25] In contrast, the microphone studied here encodes diaphragm strain directly in an integrated photonic waveguide, making the optical path itself part of the mechanical sensing element.

# III.NOISE CONTRIBUTION

Low self-noise and a high SNR are important figures of merit for modern-day commercial MEMS microphones. Generally speaking, the self-noise of a capacitive MEMS microphone mainly comes from three components: the backplate, back volume, and ASIC (aim specific integrated circuit), as shown in Figure 2.[14] Figure 2(a) shows a pie chart of noise contributors. Figure 2(b) presents a basic lumped model for a conventional capacitive MEMS microphone. Related studies of lithium niobate microphones and packaged AlN (aluminum nitride) piezoelectric directional microphones address sensitivity, noise, and the interaction between the mechanical sensor and its electrical readout.[23,26] These provide useful comparisons when assessing the noise budget of an optical microphone.

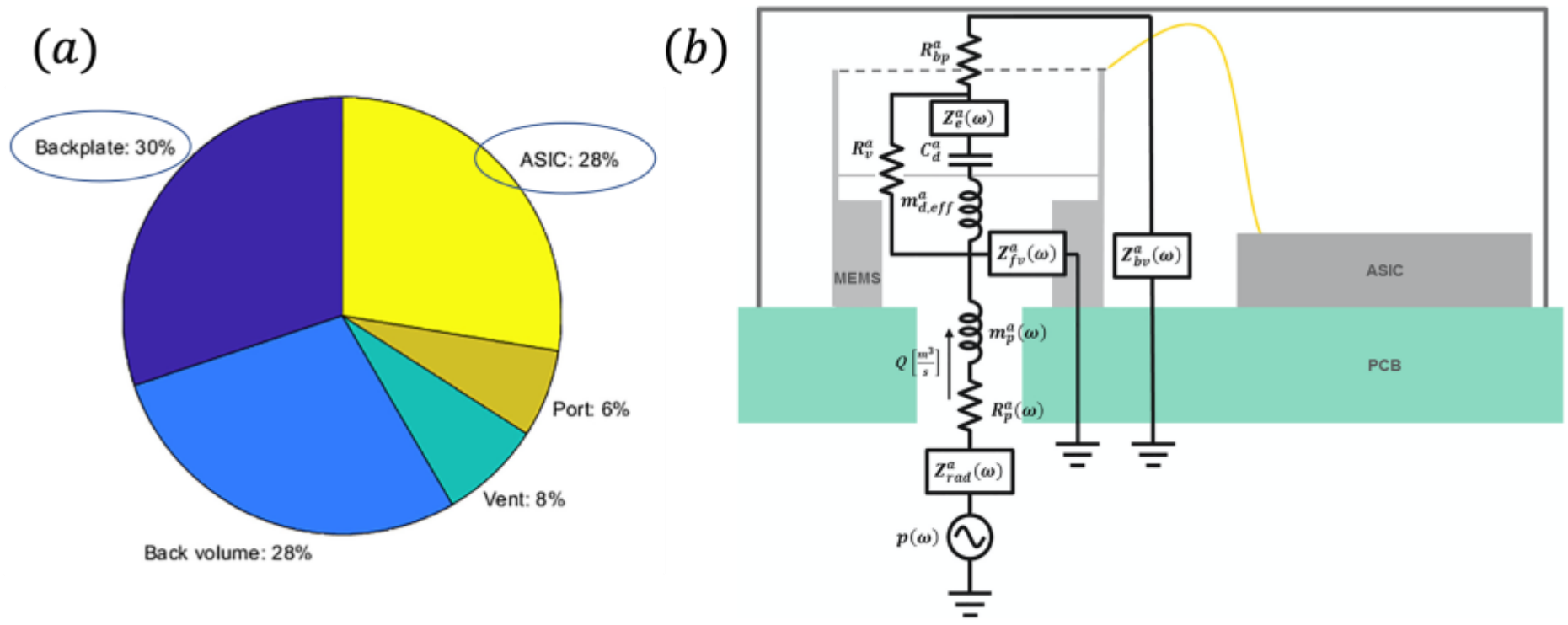

***Figure 2(a) Pie chart of noise contributors of a capacitive MEMS microphone. (b) Lumped-element model of a capacitive MEMS microphone.***

A conventional optical microphone (i.e., an optical grating microphone) typically includes a MEMS diaphragm, optoelectronics, an ASIC, a protective lid, and a PCB, as shown in the CAD illustration in Figure 3(a).[15] Figure 3(b) presents a zoomed-in image of the MEMS diaphragm and optoelectronics. A VCSEL (vertical-cavity surface-emitting laser) illuminates the optical grating. Some light is reflected by the optical grating. Some light passes through the grating and is reflected by the silicon diaphragm. If the silicon diaphragm is static, the difference between the two optical paths (i.e., the phase difference) stays the same. The optical power from the interference of these two beams will oscillate due to vibration of the silicon diaphragm. Thus, the acoustic pressure can be obtained through measurement of the interference signal by a photodiode. The grating of such an optical microphone can be considered a modified version of the backplate of a capacitive MEMS microphone. Larger holes in the grating lead to reduced thermomechanical noise. Electronic noise (i.e., ASIC noise shown in Figure 2(a)) associated with high input-impedance amplification, as required for small capacitive and piezoelectric sensors, is also circumvented. The packaged microphone on a development board is shown in Figure 3(c). With a package size of $4.14\,mm \times 4.15\,mm \times 1.38\,mm$, the optical method of transduction achieved a noise floor of $22.00\,dBA$, equivalent to a $72\text{-}dB$ SNR, as shown in Figure 3(d). Compared with common commercial capacitive MEMS microphones, the optical microphone demonstrated lower noise (i.e., a higher SNR). Note that the work summarized in Figure 3[15] was prototyped in 2014. Nowadays, state-of-the-art capacitive MEMS microphones have improved significantly. Yet, optical transduction remains an interesting method to build a MEMS microphone due to the high flexibility and integrability afforded by the development of integrated photonics technology.

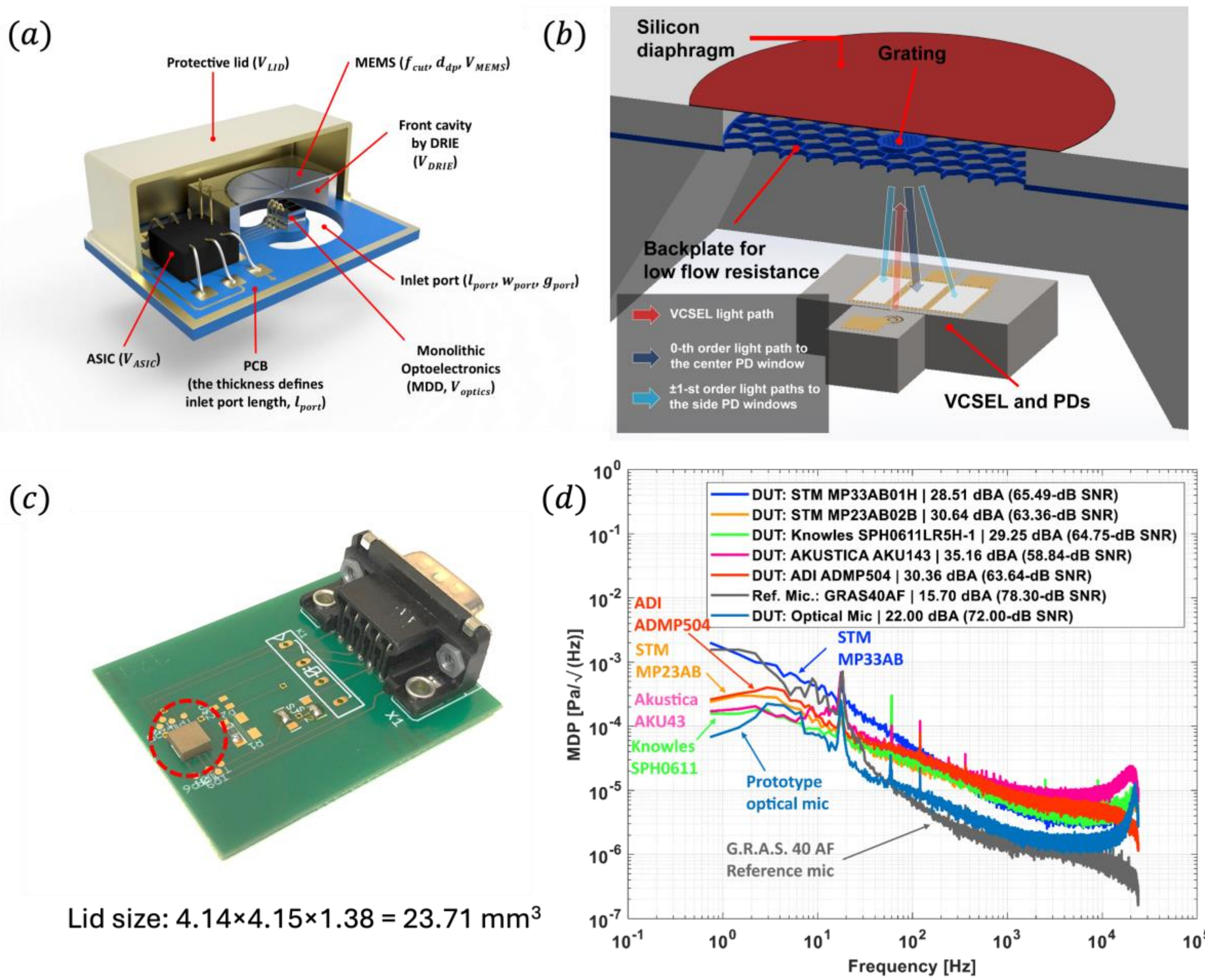

***Figure 3(a) CAD image of the design of an optical grating microphone. (b) Zoomed-in image of the MEMS die of an optical grating microphone. (c) Optical grating microphone prototyped on a test PCB. (d) MDP comparison between an optical grating microphone and other MEMS microphones on the market in 2014.***

To decrease the noise floor even further, an optical waveguide microphone uses a unique transduction scheme that does not require a backplate or grating plate because interference occurs on the diaphragm. Figure 4 presents the conceptual comparison among a capacitive MEMS microphone, an optical grating microphone, and an optical waveguide microphone. Each black block represents one kind of microphone and the associated pie chart of noise contributions. Compared to a capacitive MEMS microphone (Figure 4(a)), an optical grating microphone (Figure 4(b)) has larger holes in the grating structure, resulting in lower thermomechanical noise. Thus, the pie chart area representing backplate noise is smaller (i.e., the purple part) in the pie chart of Figure 4(b). Because such a design requires no high input-impedance amplification, optical transduction intrinsically eliminates ASIC noise. Furthermore, an optical waveguide microphone enables the elimination of both the backplate and ASIC noise in the pie chart of Figure 4(c).

The removal of a high-input-impedance electrical interface should be distinguished from the elimination of all readout noise. Our magnetoelastic vibration sensors offer another example of how transduction changes the interface requirement: their low output impedance can permit operation without local amplification electronics.[31] For the optical microphone, photodetector and subsequent electronic noise must still be evaluated alongside shot noise in a complete implementation.

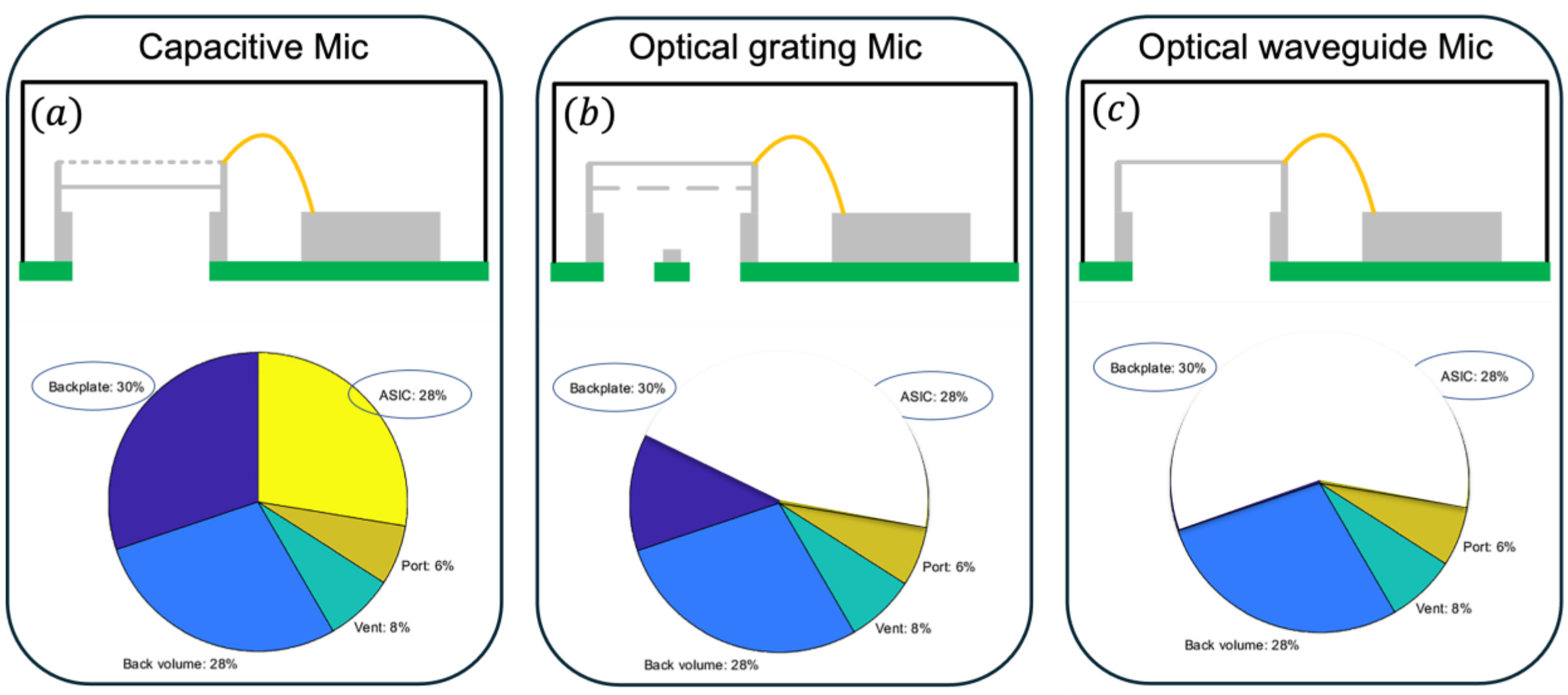

***Figure 4 Conceptual designs of three kinds of MEMS microphones and their noise contributors.***

# IV.DESIGN PROCEDURE

Figure 5 presents the basic structure of an optical waveguide microphone. As mentioned in the previous section, it is topologically equivalent to a MEMS die with no backplate, so only one diaphragm (or membrane) is shown in Figure 5(a). Other components, such as the ASIC, PCB, bonding wire, and lid, are the same as those of a regular MEMS microphone. On the MEMS die, the most important components are waveguides, as Figure 5(b) shows. The light splits and recombines via an on-chip nanofabricated MMI (multimode interferometer). After splitting, the light travels along a reference MZI (Mach–Zehnder interferometer) arm on the die substrate and a sensing MZI arm on the membrane. Acoustic pressure causes the membrane to vibrate, changing the waveguide length of the sensing MZI arm. (This mechanical process is described in detail in the next section of this article.) As a result, an optical path difference is generated. The phase difference eventually results in a change in output optical power after the light from the two optical paths is combined. Figure 5(c)[12] presents a realistic design of such an MZI. Subsequent work by our group has explored an integrated-photonic microphone with a micro resonator embedded in a diaphragm and a thermal-optic tuner.[27] That resonant architecture provides a related approach to converting diaphragm deformation into an optical signal.

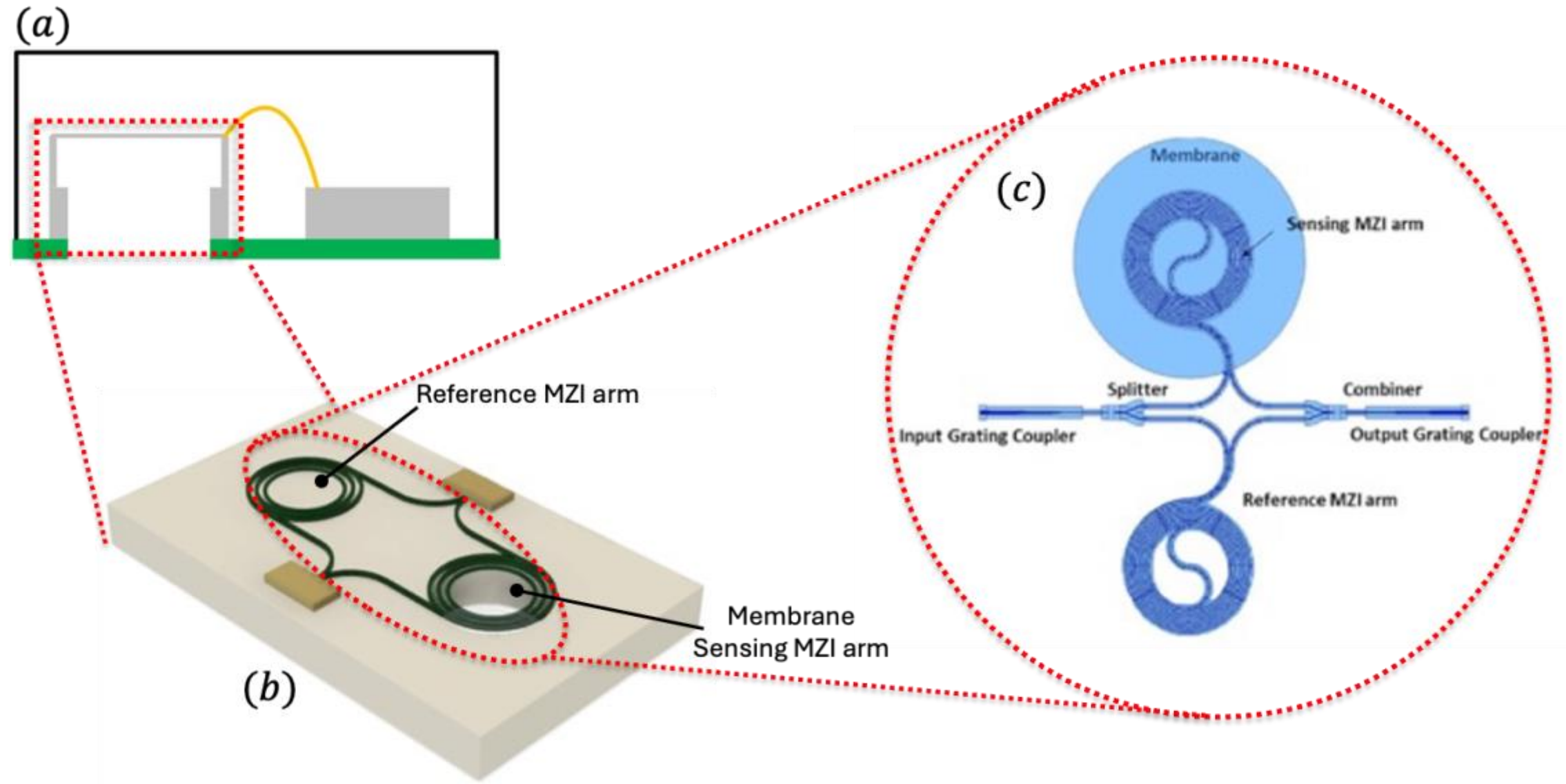

***Figure 5 (a) Conceptual design of an optical waveguide microphone. (b) Waveguides on the MEMS die, including the reference MZI arm and the sensing MZI arm on the membrane. (c) Schematic of a realistic design of MZI waveguides.***

With the development of integrated photonics, it is possible to integrate optical waveguides, a laser emitter[16], and a photodetector[17] on a single die, as shown schematically in Figure 6. Note that an edge-emitting laser is preferred for heterogeneous integration instead of a vertical-external-cavity surface-emitting laser (VECSEL). Our photonic accelerometer implementations provide related examples of interfacing waveguide sensing structures with external optical readout components.[20,21] Such arrangements offer a practical intermediate step when complete integration of the light source and detector is unavailable.

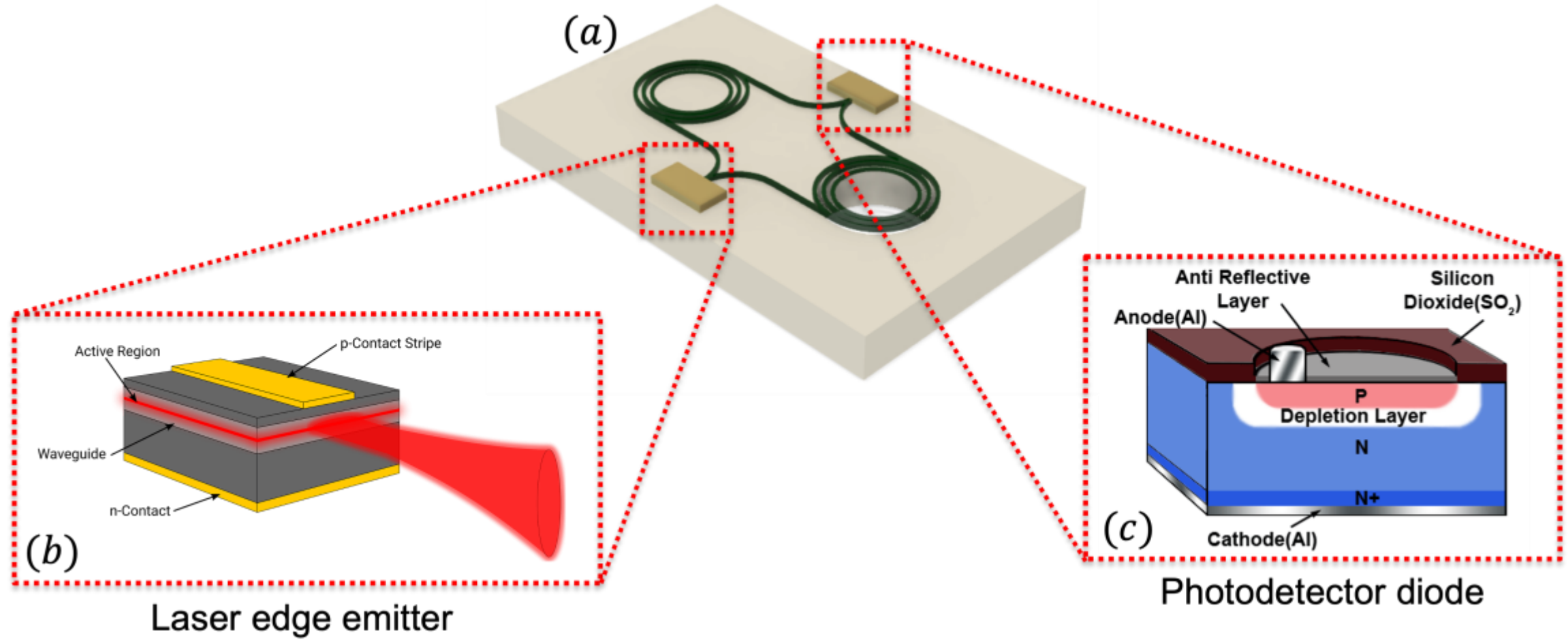

***Figure 6 (a) Conceptual design of the MEMS die of an optical waveguide microphone. (b) Laser source on the MEMS die. An edge-emitting laser is preferred. (c) Photodetector on the MEMS die.***

## A.INTRINSIC DYNAMIC RANGE

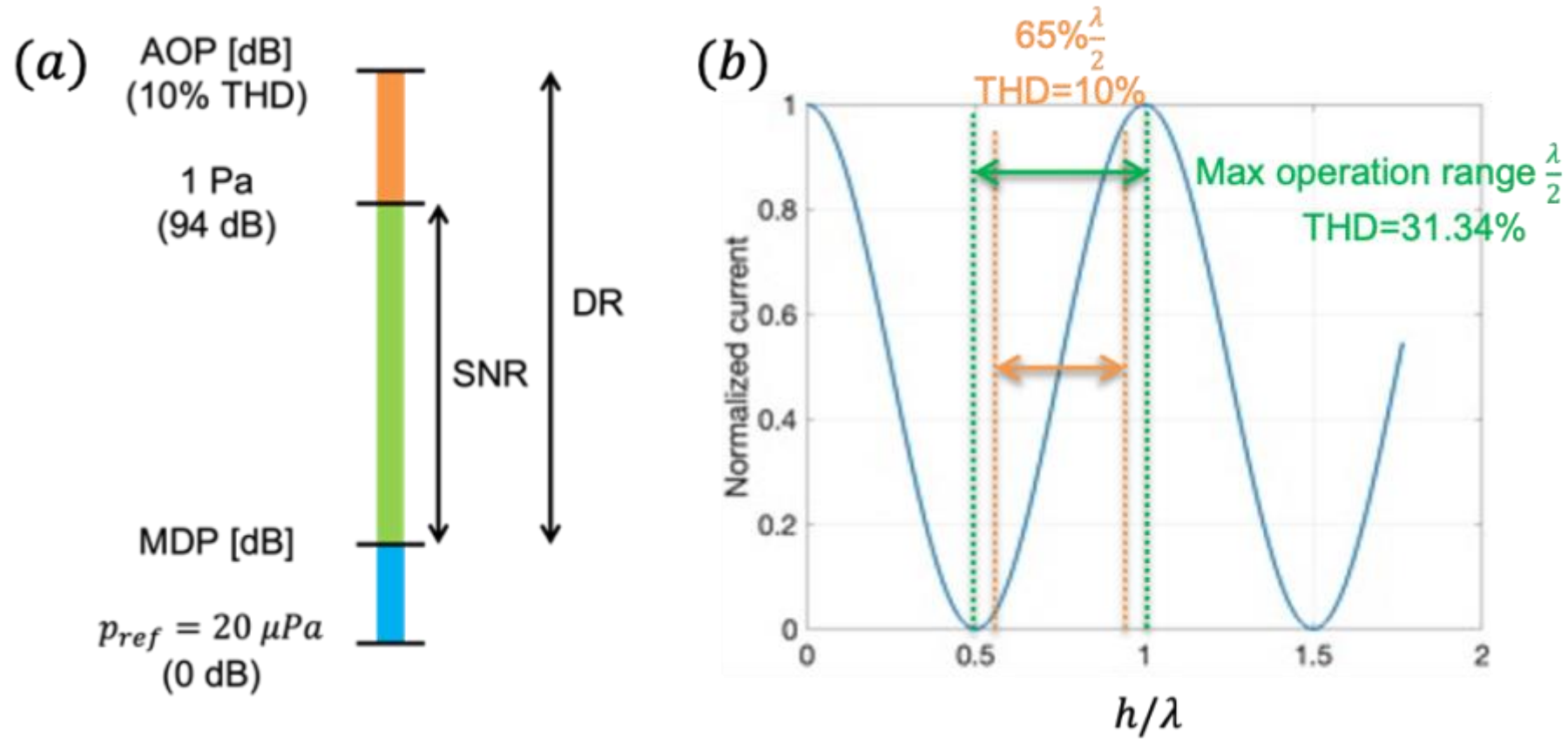

***Figure 7 (a) A color bar showing the relationship among MDP, SNR, DR, and AOP. (b) Current response of the photodetector with respect to the normalized optical path length difference.***

The key features of a MEMS microphone are the minimum detectable pressure (MDP), signal-to-noise ratio (SNR), dynamic range (DR) and acoustic overload pressure (AOP). These quantities are not completely independent. Figure 7(a) shows the correlations. Note that the AOP and DR determine the linear operating range of a microphone. Because of the optical principle of detection, the signal transduction is much more nonlinear than the mechanical component of the system. Thus, the intrinsic DR is only limited by the nonlinearity of the optical transduction. The optical nonlinearity limit considered here applies to direct intensity readout near a fixed interferometric bias point. Our subsequent silicon-nitride accelerometer study uses heterodyne detection and in-phase/quadrature demodulation to extend the measurable phase range;[21] applying such a readout to this microphone would require a separate system-level analysis.

The output optical power from the MMI (shown in Figure 5(c)) is

$$L = \frac{L_0}{2}\left[1 + \cos\left(\frac{2\pi h}{\lambda}\right)\right]\ [W] \tag{1}$$

where $L_0$ is the input optical power (i.e., the optical power received by the photodiode shown in Figure 6(c)), $h$ is the optical path length difference generated by the on-chip MZI interferometer, and $\lambda$ is the optical wavelength. The current generated by the photodiode is:

$$i = R_{op} \cdot ME \cdot L\ [A] \tag{2}$$

where $R_{op}$ is the responsivity of a photodetector in units of $\left[\frac{A}{W}\right]$, and $ME$ is the modulation efficiency in units of $[\%]$. Substituting equation 1 into equation 2, we obtain the current output as a function of optical path difference, as shown in Figure 7(b). Note that the maximum linear operating range is $\frac{\lambda}{2}$, as marked by the green dashed line. The upper limit of DR (i.e., AOP) is defined by 10% THD, which corresponds to a $65\%\frac{\lambda}{2}$ optical path difference swing.

Optical sensitivity is

$$S_{op} = \frac{di}{dh}|_{max} = \frac{\pi}{\lambda}\sqrt{ME \cdot R_{op}L_0}\ \left[\frac{A}{m}\right] \tag{3}$$

To determine the upper limit of DR, we consider only shot noise as the fundamental noise source. Thus, the minimum detectable displacement (MDD) is written as

$$MDD = \frac{i_{shot}}{S_{op}} = \frac{\sqrt{qR_{op}L_0}}{\frac{\pi}{\lambda}\sqrt{ME \cdot R_{op}L_0}} = \frac{\lambda}{\pi}\sqrt{\frac{q}{ME \cdot R_{op}L_0}}\ \left[\frac{m}{\sqrt{Hz}}\right] \tag{4}$$

Note that MDP characterizes the performance of the entire microphone system, whereas MDD describes only the optical transduction component in terms of optical path length difference. The dynamic range is then

$$DR = \frac{65\% \cdot \frac{\lambda}{2}}{MDD} = 65\% \cdot \frac{\pi}{2} \cdot \sqrt{\frac{ME \cdot R_{op} L_0}{q}} \quad (5)$$

Input-referred noise provides a common basis for comparing different transduction mechanisms. In our magneto-strictive hydrophone study, minimum detectable pressure is estimated from modeled self-noise and sensitivity.[32] A comparable assessment of the waveguide microphone should identify the frequency-dependent pressure sensitivity and measurement bandwidth and distinguish a calculated fundamental limit from an experimentally demonstrated noise floor.

The DR is usually described in units of $dB$. We note that DR is proportional to the square root of the optical power input to the waveguides. To analyze $DR$ quantitatively, we use common values for each parameter in equation 5, for example, $ME = 50\%$ or $100\%$, $R_{op} = 0.5 \left[\frac{A}{W}\right]$, and $q = 1.6 \times 10^{-19}$ $[C]$. Table 1 presents the calculated values of $DR$. To understand the results, the performance of a state-of-the-art (SOA) MEMS microphone is also listed in one column. Generally, the DR of an SOA MEMS microphone can reach 110 $dB$ with a power consumption of around 1 $mW$. For comparison, the $DR$ of an optical waveguide microphone does not offer an advantage, whether $ME = 100\%$ or $ME = 50\%$.

***Table 1 Calculated DR values for different optical input powers.***

| $L_0$[mW] | $DR[dB]$ ($ME = 100\%$) | $DR[dB]$ ($ME = 50\%$) | SOA MEMS Mic |
|---|---|---|---|
| 1 | 115 | 112 | 110dB |
| 10 | 125 | 122 | |
| 100 | 135 | 132 | |

### B.SNR AND AOP

For a microphone, SNR and AOP are both referred to acoustic pressure. The operation of an optical microphone depends on the combination of mechanical transduction and optical transduction. Mechanical transduction converts acoustic pressure into an optical path length difference. Then, optical transduction converts the optical path length difference into a photocurrent output. The principle of optical transduction is already illustrated in Figure 7. Compared to optical transduction, the mechanical transduction is much more linear. The combined transduction is equal to mechanical transduction multiplied by optical transduction. Figure 8 presents the total response of an optical microphone, which has the same shape as Figure 7(b) except for the horizontal axis. Note that the blue curve represents the static response relative to atmospheric pressure. Acoustic pressure is treated as an oscillating signal atop the bias point, i.e., the maximum slope location. As a result, the optical microphone system produces an oscillating current signal. A microphone is operated below AOP, i.e., within 65% of the maximum operating range. Thus, the oscillating acoustic input signal lies within the window marked by the orange dashed lines.

The frequency dependence of both conversion stages also matters. Our Terfenol-D inertial sensor has a native jerk response, giving its acceleration sensitivity a +6 dB/octave slope below resonance; this behavior was examined using independent measurements and finite-element modeling.[33] Although its measurand differs, it illustrates why a sensitivity specified at one frequency cannot by itself establish broadband sensor performance.

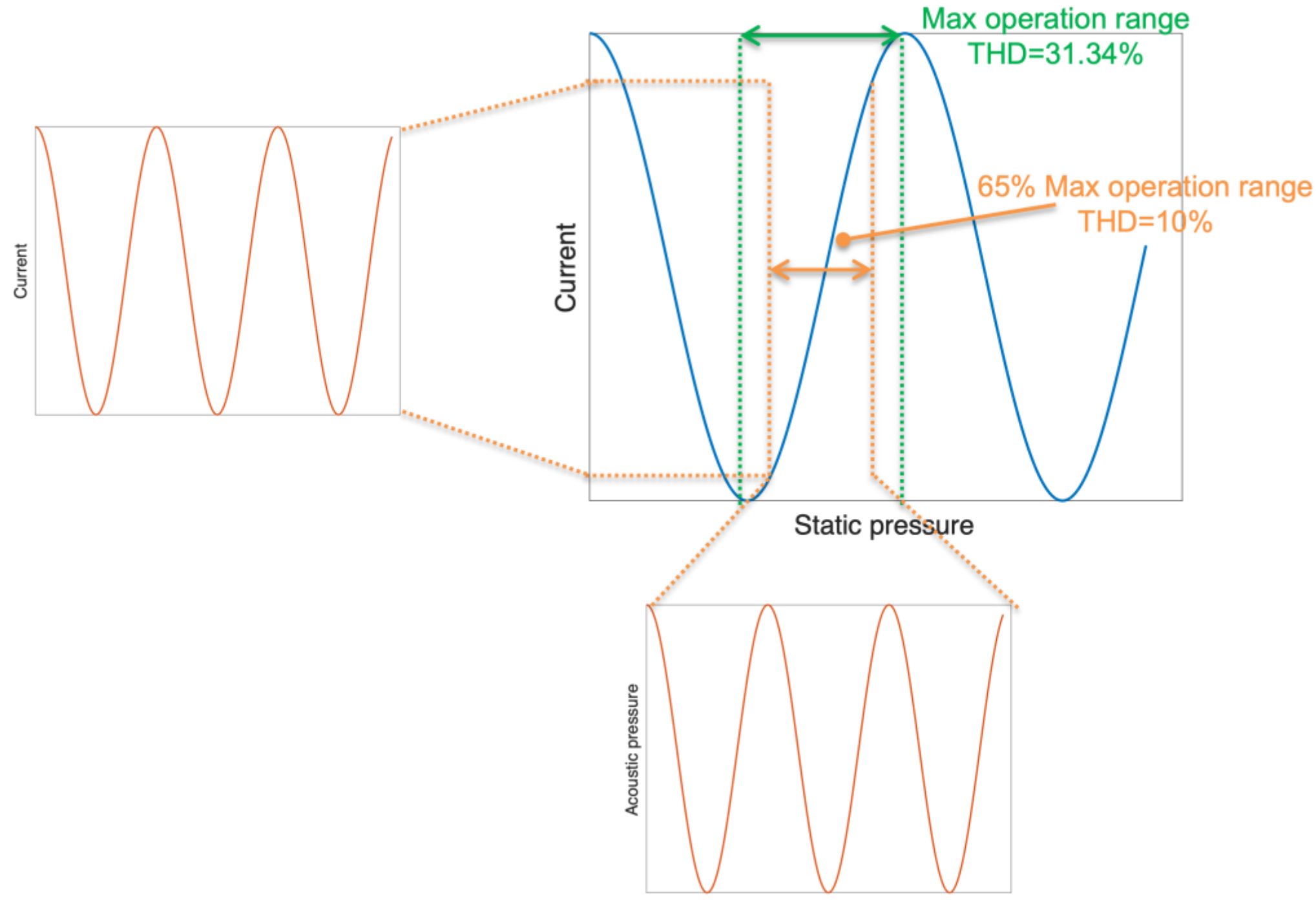

***Figure 8 Current output of the photodetector with respect to static pressure.***

For a packaged optical waveguide microphone (shown in Figure 5(a)), two major noise sources were considered, namely the quantum shot noise $i_{shot}$ and thermomechanical noise in the back cavity $i_{BC}$ in units of amperes. Thus, considering these two noise sources, the signal-to-noise ratio can be written as:

$$SNR = \frac{1[Pa] \cdot S_T \left[\frac{A}{Pa}\right]}{\sqrt{i_{shot}^2 + i_{BC}^2}[A]} \tag{6}$$

$$S_T = S_{op}\left[\frac{A}{m}\right] \cdot S_m \left[\frac{m}{Pa}\right] \tag{7}$$

where $S_T$ is the total sensitivity for an optical waveguide microphone. The total sensitivity is the product of optical sensitivity ($S_{op}$ $\left[\frac{A}{m}\right]$ derived in equation 3) and mechanical sensitivity ($S_m$ $\left[\frac{m}{Pa}\right]$). Mechanical sensitivity is described by the following equation:

$$S_m = \frac{dh}{dp}\left[\frac{m}{Pa}\right] \tag{8}$$

where $p$ is the acoustic pressure applied to the membrane of the microphone. Obtaining an expression for $h$ in terms of $p$ is not trivial. The relation is derived as follows.

A cross-sectional schematic of a MEMS die is shown in Figure 9(a). $R$ is the radius of the membrane, $th$ is the thickness of the membrane, $z$ is the horizontal distance between the center of the waveguide and the center of the membrane, and $r_n$ is the radius of the $n^{th}$ turn of the waveguide. It can be noted that in the cross-sectional schematic, there is only one waveguide on the membrane, but in practice there can be multiple turns as shown in Figure 5(c). This is due to the simplification of the schematic. The membrane is deformed by the acoustic pressure as shown in Figure 9(b). $w(r_n)$ and $\Delta r_n$ represent the resulting vertical and horizontal displacements of the $n^{th}$ turn of the waveguide, respectively. Note that the membrane expands under acoustic pressure and the waveguides are elongated. $\Delta r_n$ is the change in radius of the $n^{th}$ turn of the waveguide on the membrane. Thus, the resulting optical path length change is

$$h = \sum_{n=1}^{N} 2\pi \cdot \Delta r_n \tag{9}$$

For a total of $N$ waveguide turns, $\Delta r_n$ and $w(r_n)$ are not independent. From the geometric relations shown in Figure 10, one gets

$$\Delta r_n = -z \times \frac{dw(r_n)}{dr_n} \tag{10}$$

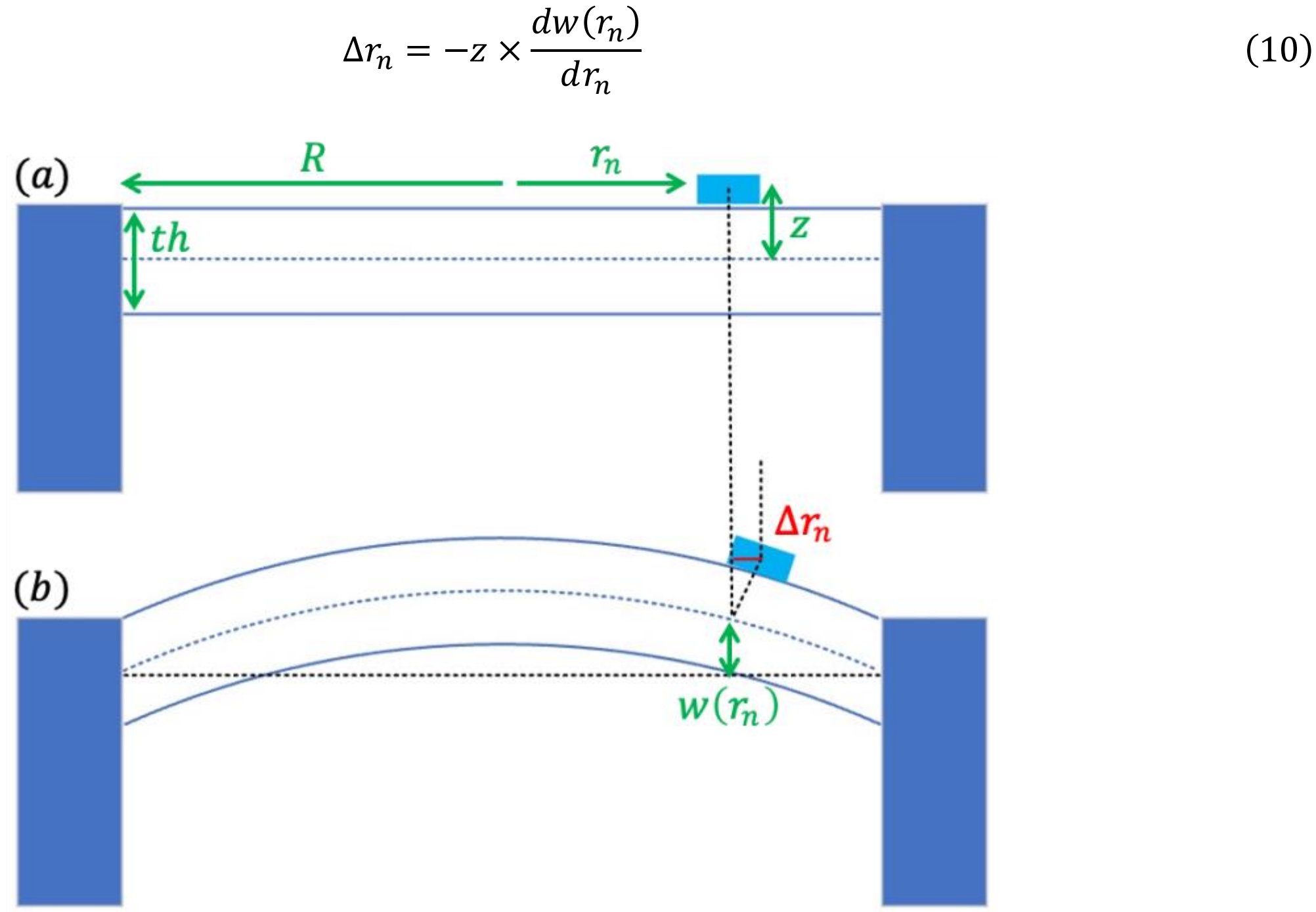

***Figure 9 Schematic cross-section of the MEMS die of an optical waveguide microphone.***

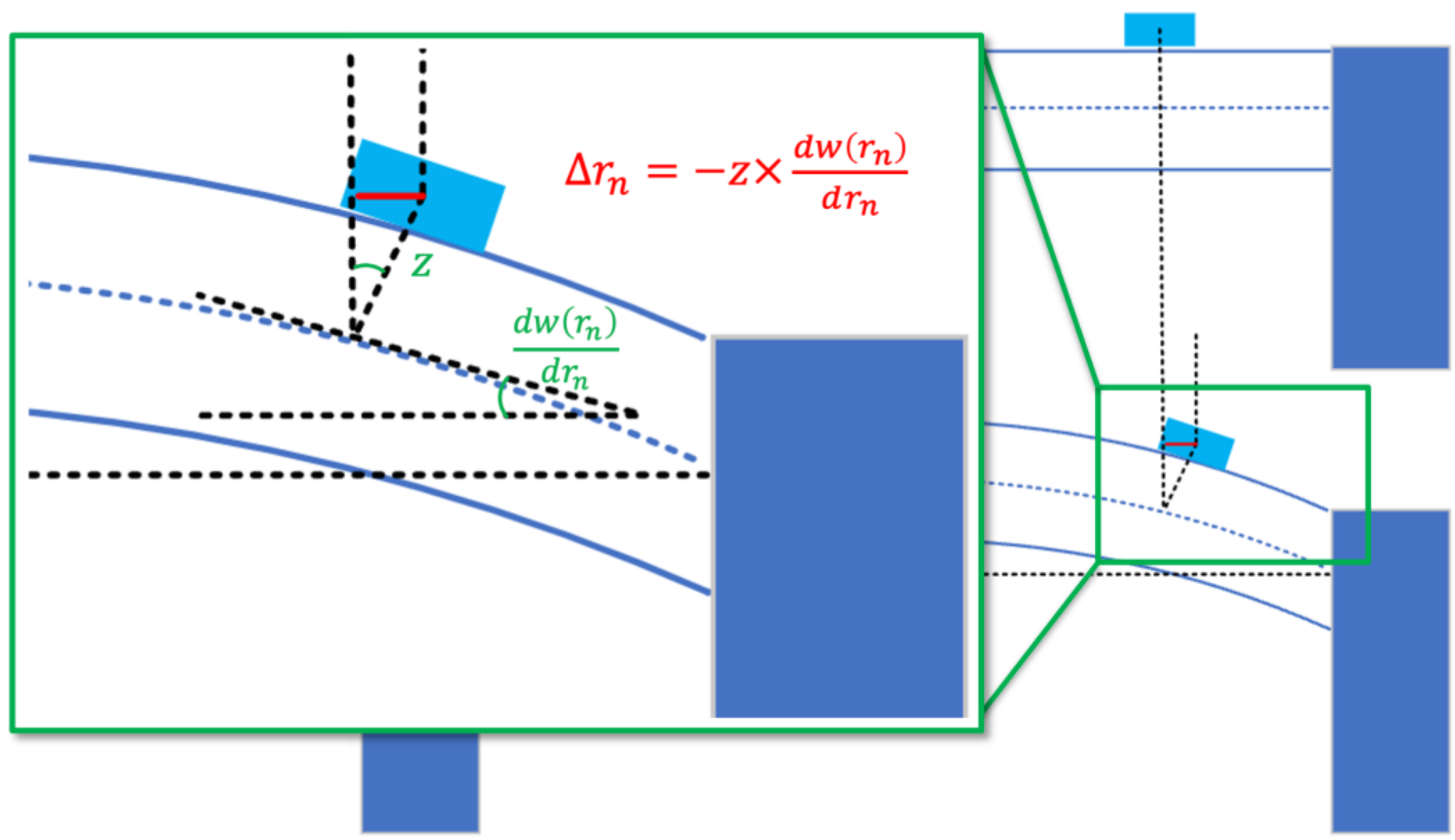

***Figure 10 Zoomed-in schematic of the interaction between an optical waveguide and the membrane.***

We consider the diaphragm deflection profile using a plate model. Based on the same parameters, the plate assumption means that bending stiffness dominates over intrinsic stress. Elongation of the waveguides results from strain on the surface of the membrane. Therefore, larger optical path length differences are produced given the same pressure. To conclude, if the diaphragm is dominated by bending stiffness, greater mechanical sensitivity will be achieved; otherwise, it will result in lower mechanical sensitivity. The plate assumption is preferred to determine the upper limit of the performance of the proposed optical microphone. Our cantilever-based photonic accelerometer uses multiple serpentine waveguide loops to increase sensitivity.[20] This suggests optimizing both the number of loops and their placement within the diaphragm strain field, rather than treating waveguide length alone as the design variable.

The material stack is another design variable. Our planar single-layer lithium niobate transducer demonstrates flexural operation without a passive layer by using thickness-shear piezoelectric coupling.[34] Together with our bimorph lithium niobate microphone,[23] this provides structural alternatives against which to consider the fabrication complexity and mechanical loading introduced by a diaphragm-integrated optical waveguide.

$$w(r_n) = w_0\left(1-\frac{r_n^2}{R^2}\right)^2 \tag{11}$$

$$w_0 = C_m \cdot p \tag{12}$$

$$C_m = \left(\frac{16E\cdot th^3}{3R^4(1-\nu^2)}+\frac{4\sigma\cdot th}{R^2}\right)^{-1} \tag{13}$$

where $w_0$ is the deformation or displacement at the center of the membrane. $C_m$ is the mechanical compliance of the microphone membrane.[18] In the expression for $C_m$, $E$, $\nu$, and $\sigma$ represent Young's modulus, Poisson's ratio, and intrinsic stress, respectively.

A fabricated device could be evaluated by measuring diaphragm motion and acoustic pressure independently before assessing the optical conversion. Our immersion lithium niobate PMUT study combines LDV and hydrophone measurements to characterize mechanical motion and acoustic output.[35] For the present airborne receiver, an analogous validation would use LDV together with a calibrated reference microphone under the same acoustic excitation.

Thus, substituting equations (10) - (12) back into equation (9) gives

$$h = \sum_{n=1}^{N}\frac{8\pi z r_n}{R^2}\left(1-\frac{r_n^2}{R^2}\right)C_m p\ [m] \tag{14}$$

$$S_m = \sum_{n=1}^{N}\frac{8\pi z r_n}{R^2}\left(1-\frac{r_n^2}{R^2}\right)C_m\ \left[\frac{m}{Pa}\right] \tag{15}$$

Then, substituting equation (3), equation (15), and $i_{shot}$ into equation (6), we obtain

$$SNR = \frac{\frac{\pi}{\lambda}\sqrt{ME\cdot R_{op}L_0}\cdot\sum_{n=1}^{N}\frac{8\pi z r_n}{R^2}\left(1-\frac{r_n^2}{R^2}\right)C_m}{\sqrt{qR_{op}L_0+i_{BC}^2}}$$

In addition,

$$Max\ operation\ range\ [Pa] = \frac{\frac{\lambda}{2}[m]}{S_m\left[\frac{m}{Pa}\right]} = \frac{\frac{\lambda}{2}}{\sum_{n=1}^{N}\frac{8\pi z r_n}{R^2}\left(1-\frac{r_n^2}{R^2}\right)C_m}[Pa] \tag{16}$$

Acoustic overload pressure can therefore be derived analytically:

$$AOP = 65\%\cdot Max\ operation\ range\ [Pa] \tag{17}$$

Experimental determination of overload should separate source distortion from sensor distortion. Our expanded study of MEMS microphones operated as ultrasonic transmitters discusses receiver nonlinearity in measurements of a parametric array.[36] This motivates independently characterizing the excitation and measurement chain when testing whether distortion in a waveguide microphone originates in the diaphragm, optical transfer function, or readout electronics.

## V.TWO DESIGN CASES

We derived key parameters of an optical waveguide microphone in the previous sections. Based on the analytical model, we explored two design cases. One is a MEMS microphone; the other is a measurement microphone. The major difference between the two types of microphones is power consumption. The MEMS microphone is targeted at consumer electronics. Power consumption is restricted, so we set the laser power to $L_0 = 1mW$. However, there is no power limit for a measurement microphone. A laser power of $L_0 = 100\ mW$ is completely acceptable. The second difference between them is thermal acoustic noise in the back cavity. Kuntzman[14] explains that such noise increasingly dominates the noise floor of a MEMS microphone as its package becomes smaller. A MEMS microphone has a smaller package than a measurement microphone. Therefore, we use 10 $dB$ equivalent input noise (EIN) as the assumed back-cavity noise level. 10 $dB$ is a state-of-the-art value for MEMS microphones on the market. Besides, we assume the waveguides do not affect the mechanical properties of the membrane. Although this is determined by the microfabrication process, this assumption helps simplify the model for exploring the upper performance limit of an optical waveguide microphone.

Selected designs for MEMS microphones and measurement microphones obtained through iterative optimization are summarized in Figure 11 and Figure 12.

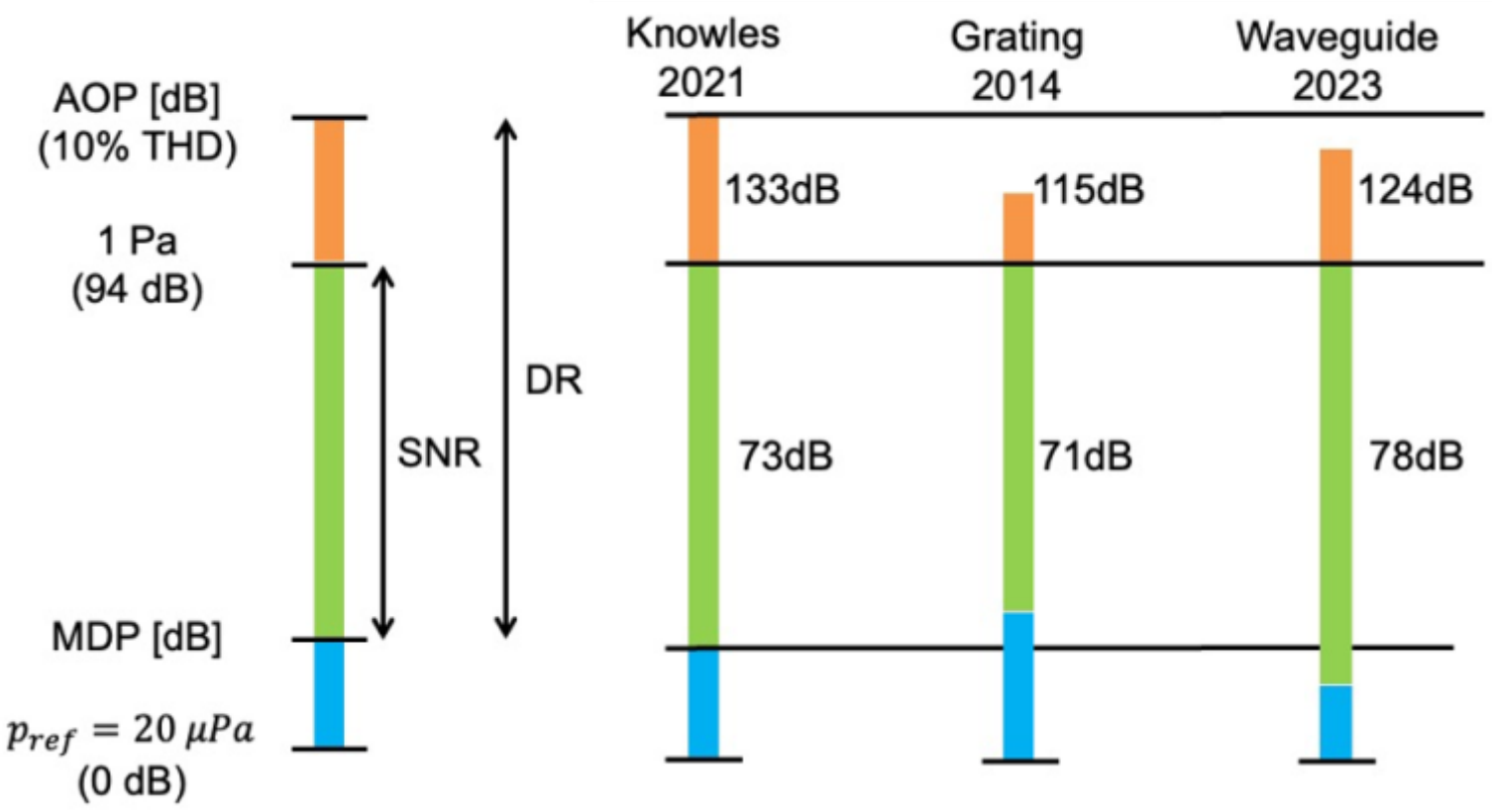

***Figure 11 An optimized optical waveguide MEMS microphone design and a performance comparison among a state-of-the-art capacitive MEMS microphone, an optical grating MEMS microphone, and an optical waveguide MEMS microphone.***

Consider a silicon membrane with a 1 $mm$ diameter, 1 $\mu m$ thickness, and 0 $MPa$ intrinsic stress. Assume a modulation efficiency of 100%. An optical waveguide MEMS microphone can reach a 78-$dB$ SNR and a 124-$dB$ AOP. Compared to the state-of-the-art MEMS microphone from Knowles in 2021[19], the optical waveguide microphone can reach a DR similar to that of the Knowles microphone, as shown in Figure 11. The waveguide microphone has advantages in SNR and AOP compared to the optical grating MEMS microphone prototyped in 2014[15]. However, the optical waveguide microphone does not show an overall performance advantage for MEMS microphone applications.

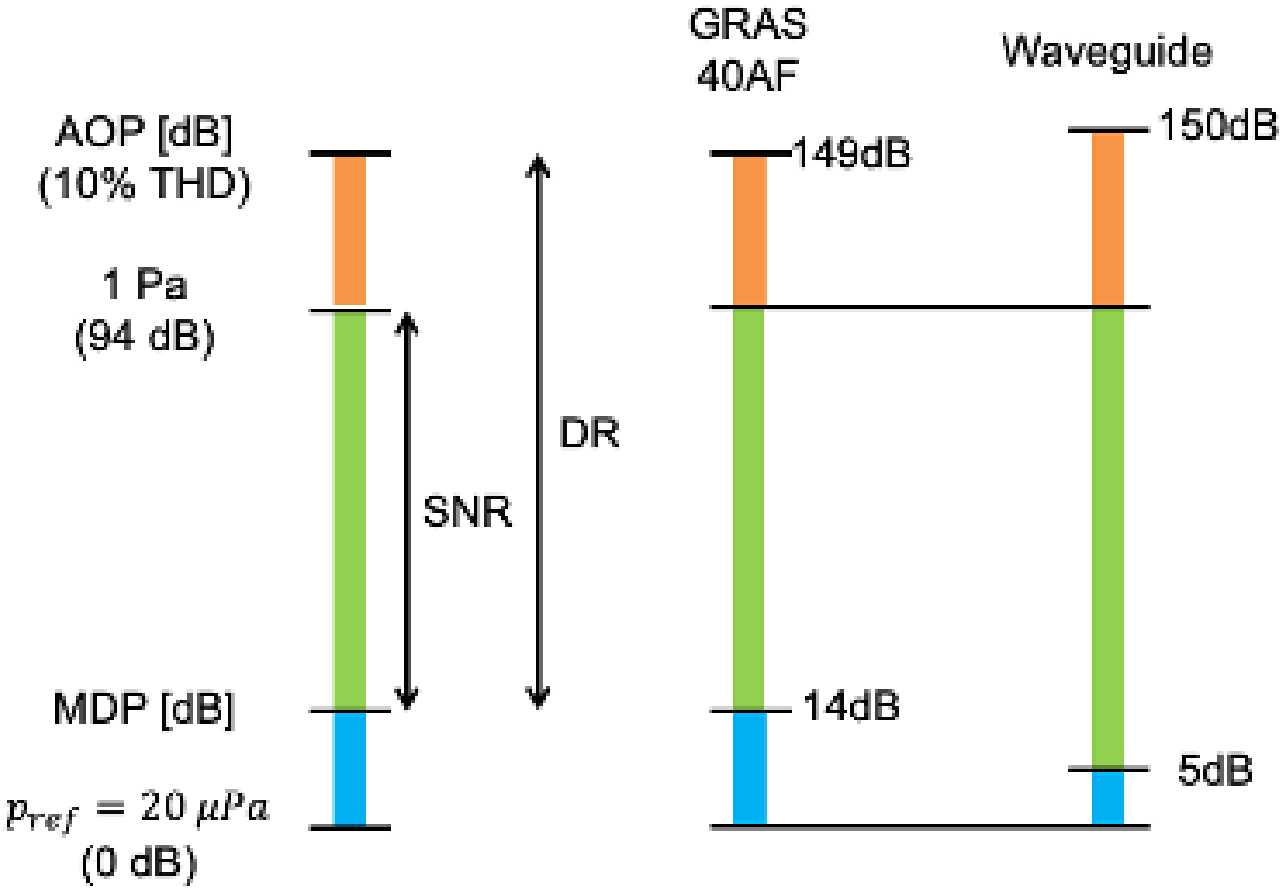

***Figure 12 An optimized optical waveguide measurement microphone design and a performance comparison between the G.R.A.S. 40AF and the optical waveguide measurement microphone.***

We use the same simulation parameters but increase the laser power to 100 $mW$ and remove back-cavity noise. It can be noted that the optical waveguide microphone can reach a lower noise floor than the G.R.A.S. 40AF measurement microphone. Although the 10 $dB$ improvement in DR is noteworthy, it mainly comes from the reduction of the MDP, so the improvement is not significant considering that the noise floor is usually not the bottleneck in measurement microphone applications. Measurement bandwidth and high-pressure waveform fidelity also matter in selecting a measurement microphone. Our study of progressive ultrasound beams at 300 kHz illustrates this requirement: resolving harmonics and shock formation required a measurement bandwidth of at least 3 MHz.[28] The present audio-band comparison does not evaluate that ultrasonic regime.

# VI.DISCUSSION

## A.MODEL INSUFFICIENCY

The model is based on several assumptions. For example, we assume that $i_{shot}$ dominates, the membrane deforms like a plate (i.e., bending stiffness dominates), and propagation attenuation in the waveguides is negligible. These assumptions intentionally yield better performance than would be expected in realistic scenarios. Thus, the analytical models presented in this article are useful for predicting the upper performance limit and conveniently estimating figures of merit. Mechanical linearity should also be checked when extending the model to large deflections. Our state-space modeling and LDV characterization of electrostatic transducers treat large diaphragm motion and associated nonlinear effects, including strain stiffening.[29] Although the actuation mechanism differs, this work motivates checking the small-deflection assumption before extending the present optical-microphone model.

Amplitude-dependent mechanical behavior can be tested with LDV. Our AlN bimorph wedge-resonator measurements identify Duffing-type nonlinearities using time- and frequency-domain data at large drive amplitudes.[37] The geometry and operating frequency differ from those of this microphone, but the measurement approach offers a way to check when the assumption of linear diaphragm motion ceases to hold.

## B.DESIGN SPACE EXPLORATION

In the case studies, we explore only the design space of the MEMS die membrane and neglect the mechanical effects of the waveguides on the membrane. Designing a MEMS microphone or a measurement microphone is much more complex than designing the membrane alone. As noted above, this simplification allows convenient analysis of the figures of merit of an optical waveguide microphone. From this perspective, the analysis in this article is reasonable. Directivity is another design objective beyond the pressure sensitivity considered here. Our fly-inspired MEMS microphone combines omnidirectional pressure sensing with two orthogonal in-plane pressure-gradient responses on a single diaphragm.[30] Extending the waveguide approach to such multimodal structures would require a different mechanical and optical design.

At the system level, the acoustic field incident on the sensor can also be engineered. Our work on coupled Helmholtz-resonator gratings demonstrates broadband, angle-dependent redirection of acoustic energy.[38] This suggests a separate avenue for exploring directional reception with an acoustic front end; the demonstrated grating is macroscopic, so compatibility with a miniature microphone would require additional scaling and loss analysis.

### C.HIGH TEMPERATURE APPLICATIONS

We have not yet mentioned another characteristic of an optical waveguide microphone. Regardless of its acoustic performance, an optical MEMS microphone can survive a high-temperature environment. We do not have to place the laser emitter, photodetector, and ASIC close to the MEMS die. Instead, we can place them at a distance from the MEMS die and use optical fibers to carry laser light to and from the die. In this configuration, the MEMS die itself becomes a high-temperature microphone because the membrane and optical waveguides are relatively stable with respect to temperature. This remarkable feature could also open the door towards supersonic or geothermal applications.

High-temperature capability should be established by measurement and should distinguish operation at temperature from survival after exposure. Our bimorph lithium niobate PMUT study reports stable operation up to 600 °C and survival up to 900 °C.[39] These results provide a useful experimental benchmark for harsh-environment acoustic transducers, but do not establish the temperature limit of the proposed optical microphone. The diaphragm, waveguide, optical coupling, package, and temperature dependence of the interferometric bias would each require characterization.

## VII.CONCLUSION

We start with the classic Michelson-Morley experiment and LIGO. A similar concept enables the measurement of acoustic pressure. The motivation for exploring an optical microphone is its low self-noise floor. We provide the basic design concept of an optical waveguide microphone including a laser emitter, MZI waveguides, a photodetector, an ASIC, and a package. To assess its figures of merit, we analyzed the upper limits of DR, SNR, and AOP for an optical waveguide microphone. Two case studies explore optical waveguide microphones for MEMS microphone and measurement microphone applications. Ultimately, we did not find substantial performance advantages for these two applications. However, high-temperature capability could offer an interesting application for an optical waveguide microphone.

## VIII.REFERENCES